# Engineering the polar magneto-optical Kerr effect in strongly strained $L1_0$-MnAl films

Lijun Zhu[1, 2*], Liane Brandt[2], and Jianhua Zhao[1]

*1. State Key Laboratory of Superlattices and Microstructures, Institute of Semiconductors,
Chinese Academy of Sciences, P. O. Box 912, Beijing 100083, China*
*2. Institut für Physik, Martin-Luther-Universität Halle, von-Danckelmann-Platz 3, Halle 06120, Germany*

**Abstract:** We report the engineering of the polar magnetooptical (MO) Kerr effect in perpendicularly magnetized $L1_0$-MnAl epitaxial films with remarkably tuned magnetization, strain, and structural disorder by varying substrate temperature ($T_s$) during molecular-beam epitaxy growth. The Kerr rotation was enhanced by a factor of up to 5 with $T_s$ increasing from 150 to 350 ºC as a direct consequence of the improvement of the magnetization. A similar remarkable tuning effect was also observed on the Kerr ellipticity and the magnitude of the complex Kerr angle, while the phase of the complex Kerr angle appears to be independent of the magnetization. The combination of the good semiconductor compatibility, the moderate coercivity of 0.3-8.2 kOe, the tunable polar MO Kerr effect of up to ~0.034º, and giant spin precession frequencies of up to ~180 GHz makes $L1_0$-MnAl films a very interesting MO material. Our results give insights into both the microscopic mechanisms of the MO Kerr effect in $L1_0$-MnAl alloys and their scientific and technological application potential in the emerging spintronics and ultrafast MO modulators.
**Keywords:** Kerr effect, Magnetic anisotropy, Spintronics, Magnetooptical modulator

\* Author to whom any correspondence should be addressed. Email: zhulijun0@gmail.com

## 1. Introduction

In recent few years, the polar magneto-optical (MO) Kerr effect in magnetic materials, especially those with strong perpendicular anisotropy, is playing a crucial role in the undergoing boost of spintronics as a powerful tool in locally probing the magnetic properties and electronic structures of materials [1-7]. For example, a variety of new exciting spintronic phenomena including Snell's law for spin waves [1], direct and inverse spin Hall effect [2,3], spin orbit torques [4], magnetic skyrmion bubbles [5], magnetic vortex dynamics [6], magnetic domain wall motion [7], and terahertz frequency magnetization precession [8] have been experimentally observed taking advantage of the polar MO Kerr effect. From the viewpoint of technological applications, high-speed optic communications require optical modulators to work at very high frequencies. The polar MO Kerr effect can be used to modulate the optical polarization/intensity at spin precession frequencies [8-11]. However, conventional ferromagnetic materials, e.g. Fe [9] or Permalloy [1,10], usually have very weak magnetic anisotropy, limiting their spin precession frequencies, i.e. modulation bandwidth, to a few GHz or less. Perpendicularly magnetized MO materials with a magnetic anisotropic field ($H_k$) of ~100 kOe, e.g. $D0_{22}$-Mn$_3$Ge [8] and $L1_0$-Mn$_x$Ga [11,12], are good candidates for high-frequency MO modulators due to their hundreds-GHz frequency spin precession. Moreover, magnetic materials compatible with semiconductors in epitaxy are much advantageous due to the allowance of direct integration of MO functional devices with underlying electronic circuits. For both the spintronic and modulator applications [2-8,11], it is, therefore, of great interest to develop new kinds of perpendicularly magnetized MO materials that simultaneously have large saturation Kerr rotation ($\theta_K$), giant $H_k$, and good semiconductor compatibility.

On the other hand, despite the great theoretical efforts in past decades [13-17], the understanding of the microscopic mechanisms of the MO Kerr effect is not complete yet. Microscopically, the MO Kerr effect arises from the spin-orbit interaction (SOI) and the magnetization ($M_s$). The Kerr angles were calculated to scale in proportion to SOI strength ($\xi$)[14]. As for $M_s$, the Kerr effect was proposed to scale proportionally to the magnetization by off-diagonal conductivity models [15]. However, some detailed theoretical calculations [16] and experiments on Fe, Co and Ni and their alloys with Si or Ge [18] showed a rather complicated influence of $M_s$ on the Kerr angles: enlarging $M_s$ can either enhance or diminish the spectral peaks of Kerr effect. Some Heusler alloys even exhibit no correlation between magnetization and Kerr angles [19]. Oppeneer *et al.* predicted that the variation of the lattice constant (i.e. strain) can remarkably shift the spectral peaks of Kerr effect in Ni and thereby significantly influence the measured Kerr angles at a fixed wavelength ($\lambda$) [16], experimental tests of which have, however, been lacking to date. Recently, we observed a strong variation of Kerr angles in a series of $L1_0$-Mn$_x$Ga ($0.76 \leq x \leq 1.29$) films with different lattice constants [11], whereas the coexistence of other composition effects, especially the significant variation in $\xi$, preventing an unambiguous conclusion with respect to whether or not the strain effect is really important. In ferromagnetic alloys, the structural disorder usually has a remarkable influence on their magnetism [20,21], Fermi surface topology [22], and electron scattering processes [22-24], which is highly reminiscent of a possible effect on MO properties. However, so far, it has remained unsettled how the structural disorder influences the MO Kerr effect in such alloy materials.

In past two decades, the noble-metal-free and rare-earth-free $L1_0$-ordered (so-called $\tau$-phased) Mn$_x$Al (MnAl) alloys have attracted increasing attention because their giant perpendicular magnetocrystalline anisotropy of ~10 Merg cm$^{-3}$, high coercivity of a few kOe, and good semiconductor compatibility make them potentially useful in ultrahigh-density perpendicular magnetic recording, spin transfer torque memories, and economical permanent magnet applications [25,26]. The magnetic and transport properties are found to vary profoundly with thickness [27], composition [21,28], strain, and structural disorder [23-26]. However, detailed accurate knowledge on the MO Kerr effect and its microscopic mechanisms in $L1_0$-MnAl epitaxial films have been still missing. Cheeks et al., reported a $\theta_K$ of 0.11° in a multilayer of GaAs (001)/AlAs 2 nm/MnAl (x was not determined) 10 nm/GaAs 20 nm at $\lambda$ = 220-820 nm [29], and $\theta_K$ of 0.06°, 0.17°, and 0.16° in multilayers of GaAs (001)/AlAs 10 nm/MnAl 10 nm/GaAs 20 nm with x = 0.67, 1.0, 1.5, respectively, at $\lambda$ = 500-800 nm [30]. Notably, the GaAs and AlAs layers surrounding the very thin MnAl layers would make the light exponentially damped and shifted in phase due to their large complex refractive index (N) in the visible light regime, e.g. $N_{GaAs}$ = 4.1 - i 0.4 and $N_{AlAs}$ = 3.38 – i 0.04 at 623.8 nm [31,32]; the interference effect within the multilayers can be probably overwhelming over the intrinsic contribution of the ferromagnetic layer [33]. Therefore, these Kerr angles reported in previous samples could be problematic and not intrinsic to $L1_0$-MnAl. It should also be pointed out that a comprehensive understanding of an MO material needs both knowledge on Kerr rotation and ellipticity ($\varepsilon_K$) or the magnitude ($|\psi|$) and phase ($\alpha$) of the complex Kerr angle $\psi = |\psi|e^{i\alpha} = \theta_K + i \varepsilon_K$. Unfortunately, there have been, so far, no report on $\varepsilon_K$, $|\psi|$, or $\alpha$ for this material. From the viewpoint of technological application, to date, there has been no relevance of $L1_0$-MnAl to possible MO applications in the emerging spintronics studies and hundreds-GHz optical polarization/magnitude modulators.

In this paper, we, for the first time, systematically studied the polar MO Kerr effect in a series of relatively thick $L1_0$-MnAl epitaxial films without any uncertainty from $\xi$ and capping layer effects. The scaling analysis revealed that the MO Kerr effect in $L1_0$-MnAl films are dominated by the magnetization as indicated by the excellent linear $M_s$-dependence of $\theta_K$, $\varepsilon_K$, and $|\psi|$, despite the considerable variations in the strain or structural disorder. These $L1_0$-MnAl films would be potentially useful in spintronic and ultrafast modulator applications due to their good semiconductor compatibility, moderate coercivity, tunable Kerr effect, and giant spin precession frequencies.

**2. Experimental methods**

A series of 30 nm-thick $L1_0$-MnAl single-crystalline films were deposited on 150 nm GaAs-buffered semi-insulating GaAs (001) substrates at a growth rate of 1 nm/min by molecular-beam epitaxy (MBE) (see Fig. 1(a)). The Mn and Al atomic ratio x was determined to be 1.1 by electron dispersive x-ray spectroscopy. The substrate temperatures ($T_s$) were varied to be 150, 200, 250, 300, 350, and 400 °C, respectively, in order to tailor their MO properties, magnetization, strain, and structural disorder [21,23]. Each film was capped with a 4 nm $Al_2O_3$ layer to prevent oxidation after cooled down to room temperature. The good homogeneity of these films are confirmed by the bright streaky reflection high-energy electron diffraction patterns, the oscillatory x-ray reflectivity curves, and the cross-sectional tunneling electron microscopy images [26]. From the atomic force microscopy (AFM) images shown in Figs. 1(b)-1(g), the RMS roughness of the $L1_0$-MnAl films was determined to be 1.9, 1.6, 1.2, 0.7, 0.5, and 0.9 nm for $T_s$=150, 200, 250, 300, 350 and 400 °C, respectively, the change tendency of which is opposite to that of $M_s$ (see Table I). The good homogeneity and relatively small roughness of these films indicate a considerably small spatial variation of surface properties and the Kerr signals. The polar MO Kerr effect measurements were carried out at room temperature by an optical setup with two laser diodes at $\lambda$=400 nm and 670 nm, respectively (see Ref. [11] for more details). Note that the thickness of MnAl layers is safely larger than the penetration length of the detection light for the MO properties (typically ~10-20 nm in ferromagnetic metals) and the imaginary component of the refractive index of $Al_2O_3$ is negligibly small in the visible light regime ($N_{Al2O3}$ = ~1.6 + i 0)[34]. Therefore, the measured MO properties should be intrinsic to the MnAl layers as they are influenced neither by the thin capping layers nor buffer layers. In order to evaluate the strain, the disorder and $M_s$ of these $L1_0$-MnAl films, the out-of-plane lattice constant (c), the static defect scattering induced residual resistivity ($\rho_{xx0}$), and the magnetic properties were determined by x-ray diffraction $\theta$-$2\theta$ scans at BL14B1 beamline of Shanghai Synchrotron Radiation Facility (SSRF) in China, a Quantum Design physical property measurement system (PPMS-9), and a Quantum Design superconducting quantum interference device (SQUID-5) magnetometer, respectively.

**3. Results and discussion**

Figures 2(a) and 2(b) show the hysteresis loops of polar Kerr rotation ($\theta$) and ellipticity (-$\varepsilon$) at $\lambda$ = 400 nm for $L1_0$-MnAl films with different $T_s$. These well-defined hysteresis loops indicate the perpendicular anisotropy in these films. The coercivities for rotation and ellipticity hysteresis vary between 0.3 to 8.2 kOe as $T_s$ increases from 150 to 400 °C. Such a change tendency agrees with that of switching fields in magnetization and anomalous Hall resistivity measurements [21,24]. In analogue to $L1_0$-$Mn_xGa$ films, each film shows a positive $\theta_K$ but a negative $\varepsilon_K$. The same features hold for the hysteresis loops at $\lambda$ = 670 nm.

Figure 3(a) summarizes $\theta_K$ in $L1_0$-MnAl films as a function of $T_s$. With increasing $T_s$, $\theta_K$ climbs up quickly from 0.007° (0.004°) to the maximum value of ~0.034° (0.028°), and then drops to 0.025° (0.020°) for $\lambda$ = 400 (670) nm. Similarly, as $T_s$ increases, $\varepsilon_K$ varies monotonically from -0.004° (-0.002°) to -0.037° (-0.021°) and then changes to -0.028° (-0.020°) (see Fig. 3(b)); $|\psi|$ increases from 0.008 (0.005) to 0.050 (0.0035) and then decreases to 0.038 (0.028) (see Fig. 3(c)). All the maximum values for $\theta_K$, $\varepsilon_K$, and $|\psi|$ are found at $T_s$ = 350 °C, suggesting that 350 °C is the most optimized temperature for the epitaxial growth of $L1_0$-MnAl films on GaAs substrates from the viewpoint of MO applications. The qualitatively analogous dependence of $\theta_K$, $\varepsilon_K$ and $|\psi|$ on $T_s$ suggests that they are dominated by the same mechanism(s).

In order to shed light on the microscopic mechanisms of the MO Kerr effect in $L1_0$-MnAl, in the following we discuss in detail how the Kerr angles ($\theta_K$, $\varepsilon_K$, and $|\psi|$) scale with $\xi$, $M_s$, strain, and disorder, respectively. Ab initio and relativistic band structure calculations show that $\theta_K$ scales linearly with $\xi$ [13-15]. For $L1_0$-MnAl films, $\xi$ is ~40 meV for Mn (3d) and ~20 meV for Al (3p), respectively [35]. As indicated by recent experimental and theoretical calculations on the anomalous Hall effect, Gilbert damping, and the MO Kerr effect [36-38], $\xi$ varies with sample compositions but keeps constant for samples with the same composition. Therefore, we expect a constant $\xi$ for the $L1_0$-MnAl films with different $T_s$ due to their same composition, i.e. $x = 1.1$, excluding the possibility of $\xi$ being relevant for the observed strong $T_s$-dependence of $\theta_K$, $\varepsilon_K$ and $|\psi|$. As shown in Table. I, $M_s$ increases from 22.1 to 263.4 emu cm$^{-3}$ as $T_s$ increases from 150 to 300 $^o$C, while drops slightly to 195.7 emu cm$^{-3}$ at 400 $^o$C; while $c$ and $\rho_{xx0}$ show an opposite tendency, suggesting remarkable changes in magnetism, overall strain, and structural disorder. It should also be pointed out that the values of $c$ in these $L1_0$-MnAl films are much smaller than the bulk value $c_0$ of 3.570 Å [39], revealing the strong strain that exists in these films. All these $L1_0$-MnAl films exhibit very large $\rho_{xx0}$ of 140.3-205.7 μΩ cm in comparison to highly ordered 3$d$ ferromagnets, e.g. ~4 μΩ cm for Fe film [40] and ~30 μΩ cm for $L1_0$-Mn$_{1.5}$Ga films [22], respectively, indicating the strong structural disorder in these $L1_0$-MnAl films. In order to clarify the roles of the magnetization, the strain, and the structural disorder in determining the MO Kerr effect, we further plotted $\theta_K$ and $\varepsilon_K$ at $\lambda$=400 nm in Figs. 4(a)-4(c) as a function of $M_s$, $(c_0-c)/c_0$, and $\rho_{xx0}$, respectively. As indicated by the best linear fits in Fig. 4(a), both $\theta_K$ and $\varepsilon_K$ show proportionally linear relation with $M_s$, strongly suggesting that $M_s$ dominates $T_s$-dependence of the MO properties of these films. In other words, the slopes $C_1 = d\theta_K / dM_s$ and $C_2 = d\varepsilon_K / dM_s$ are constants ($C_1 = (1.31 \pm 0.07) \times 10^{-4}$ $^o$ emu$^{-1}$ cm$^3$, $C_2 = (1.39 \pm 0.02) \times 10^{-4}$ $^o$ emu$^{-1}$ cm$^3$) for the $L1_0$-MnAl films with different levels of strain and structural disorder, which indicates that the variation of $\theta_K$ and $\varepsilon_K$ with $(c_0-c)/c_0$ (Fig. 4(b)) and $\rho_{xx0}$ (Fig. 4(c)) only reflects the fact of the magnetization being a function of the structural strain and disorder [21,23]. The same feature was found for $\lambda = 670$ nm as well. Therefore, we can safely conclude that the $T_s$-dependence of the MO properties of the $L1_0$-MnAl films is mainly determined by the variation of magnetization. It should be pointed out that the observed proportional relation between $M_s$ and Kerr angles in $L1_0$-MnAl is not universal for all magnetic materials as suggested by the complicated behaviors in Fe, Co, Ni alloys and some Heusler alloys [18,19]. Here the absence of a strong direct strain effect that was expected to shift the spectral peak of MO effect of Ni [14] may be attributed to a relatively weak wavelength dependence of Kerr angles of $L1_0$-MnAl [30]. Accordingly, the variation of $|\psi|$ in Fig. 3(c) should also be attributed to the change of $M_s$. Notably, the phase of the complex Kerr angle, $\alpha$, is ~43.3$^o$ at 400 nm as suggested by the two linear fits which are very close to each other (see Fig. 4(a)). Here, the relatively small $\theta_K$ in comparison to that reported in Ref. [29,30] may be partly due to the smaller $M_s$ of our samples and partly due to the absence of capping layer and underlayer effects [33]. It is quite plausible to significantly enhance Kerr angles ($\theta_K$, $\varepsilon_K$ and $|\psi|$) by improving $M_s$ from present maximum of 263.4 emu cm$^{-3}$ towards its theoretical value of 810 emu cm$^{-3}$ [25,35] through post-growth annealing (see Ref. [11, 41] for an example on $L1_0$-Mn$_x$Ga) because these Kerr angles scales linearly only with $M_s$ in the case of a constant $\xi$. It is also possible to enhance the Kerr effect via increasing SOI of the films, e.g. by increasing the Mn concentration ($\xi$ is larger in Mn than Al) or doping some heavy atoms, like Bi [42].

Finally, we want to mention that, in addition to the large MO Kerr effect, a giant internal magnetic anisotropic field ($H_k$) is highly desirable for high-frequency MO modulator application. In $L1_0$-MnAl, $H_k$ arising from the tetragonal lattice distortion ($c$ is shorter than inplane lattice constant $a$) is given by $H_k = 2K_u/M_s$, where $K_u$ is the uniaxial magnetocrystalline anisotropy constant which can be deduced from the area enclosed between the magnetization curves with applied fields parallel and perpendicular to the film plane [20]. As shown in Table I, the value of $K_u$ of these $L1_0$-MnAl films varies between 0.004 and 7.73 Merg cm$^{-3}$ as $T_s$ increases from 200 to 400 $^o$C, The spin precession frequency ($f$) due to the uniaxial magnetocrystalline anisotropy can be estimated at the first rough approximation following $f = g\mu_B H_k/h$, where $g$, $\mu_B$, and $h$ are the $g$-factor (~2.0), Bohr magnetron, and Plank constant, respectively. As is shown in Fig. 5, $H_k$ increases from 0.3 to 58.4 kOe as $x$ increases from 200 to 350 $^o$C, and drops to 48.3 kOe at 400 $^o$C, the tendency of which is well consistent with that of $M_s$ and $(c_0-c)/c_0$. For the most optimized temperature $T_s$=350 $^o$C, the giant $H_k$ of up to 58.4 kOe should give rise to a spin precession at ~180 GHz, allowing for ultrafast optical modulation. Experiments taking advantage of Brillouin light scattering and time-resolved MO Kerr effect with a terahertz optical pump would be interesting in the future to directly prove the hundreds-GHz frequency spin precession expected in this MO material.

### 3. Conclusion

We have presented the variable polar MO Kerr effect in perpendicularly magnetized $L1_0$-MnAl epitaxial films with remarkably tuned magnetization, strain, and structural disorder by varying growth temperature. The Kerr rotation is enhanced from 0.007$^o$ (0.004$^o$) to the maximum value of ~0.034$^o$ (0.028$^o$), and then drops to 0.025$^o$ (0.020$^o$) for $\lambda = 400$ (670) nm by increasing growth temperature from 150 to 400 $^o$C, which is mainly attributed to the variation of magnetization instead of spin-orbit coupling strength, overall strain or structural disorder degree. A significant $T_s$-tuning effect was observed on the magnitude of the complex Kerr angle, whereas the phase of the complex Kerr angle appears to be independent of $M_s$. For the most optimized $T_s$ of 350 $^o$C, the good semiconductor compatibility, the moderate coercivity of ~8.2 kOe, the large polar MO Kerr effect of ~0.034$^o$, giant spin precession frequencies of ~180 GHz, and the remarkable tunability together make $L1_0$-MnAl films a very interesting MO material. Our results not only give an insight into the microscopic mechanisms of the MO Kerr effect of $L1_0$-MnAl alloys, but also be helpful for the potential MO applications of this material in the high-sensitivity and high magnetic-noise-immunity studies of the emerging spintronic phenomena and in ~hundreds GHz-frequency MO modulators for ultrafast optic communications.

### Acknowledgments

We acknowledge the financial support by NSFC (Grant No. 61334006) and MOST of China (Grant No. 2015CB921503).

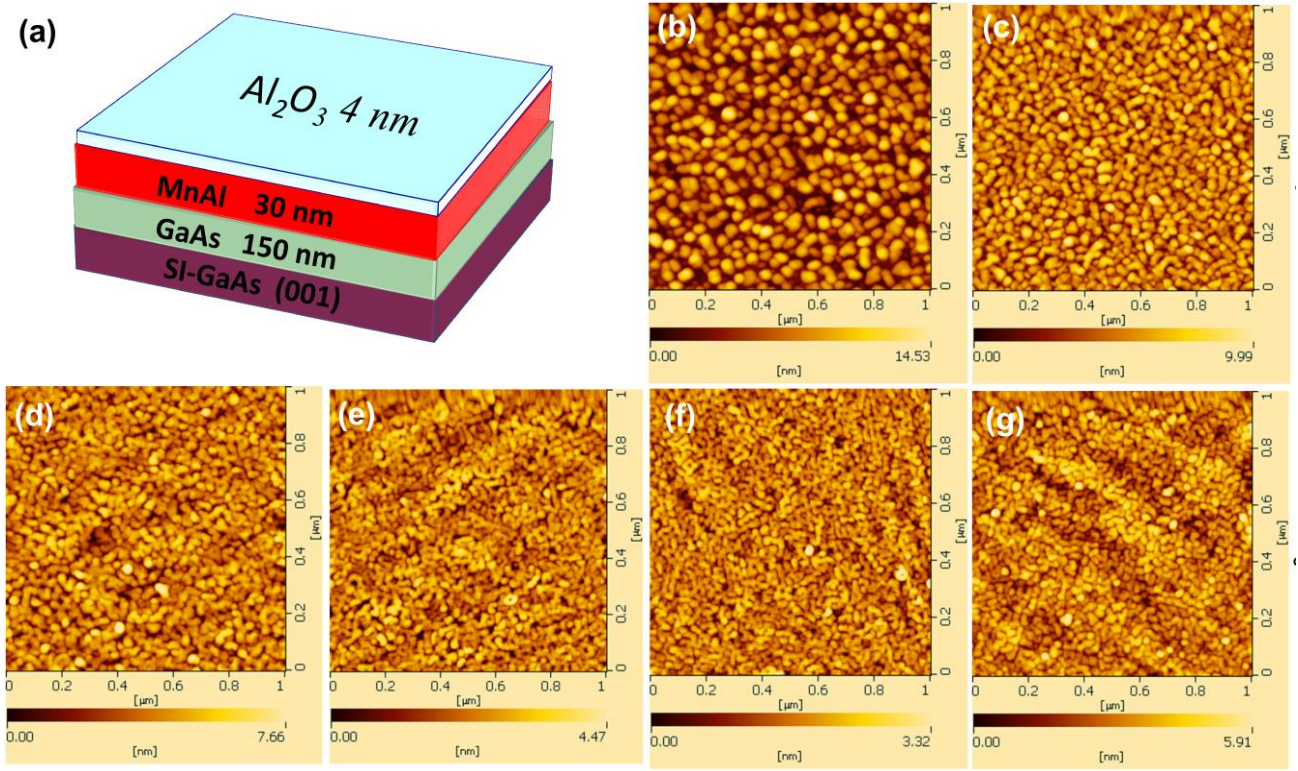

FIG.1 (a) Schematic of the sample structure; Atomic force microscopy images of $L1_0$-MnAl films grown at (b) 150 ºC, (c) 200 ºC, (d) 250 ºC, (e) 300 ºC, (f) 350 ºC, and (g) 400 ºC, respectively.

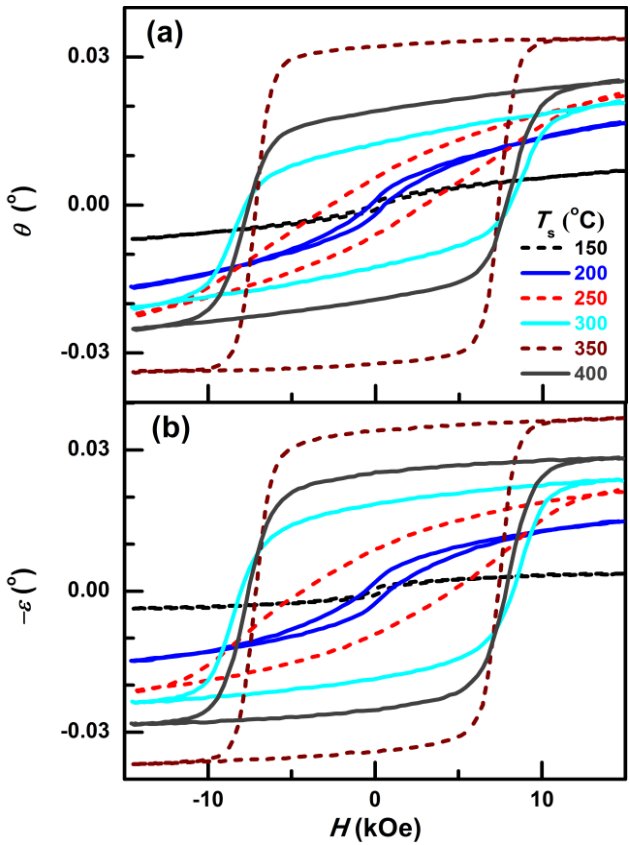

FIG. 2 Hysteresis loops of (a) Kerr rotation ($\theta$) and (b) Kerr ellipticity ($-\varepsilon$) at 400 nm for $L1_0$-MnAl films with different $T_s$.

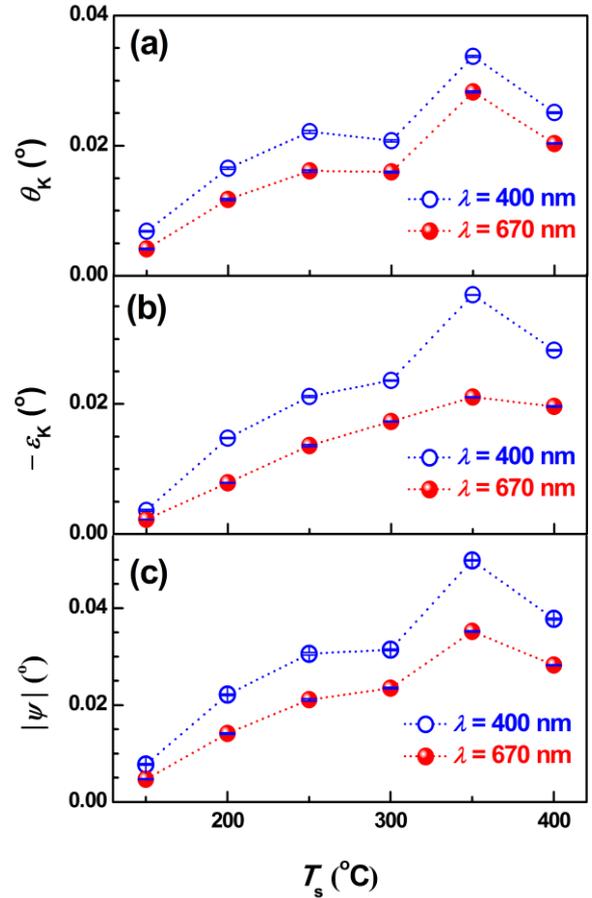

FIG. 3 $T_s$-dependence of (a) $\theta_K$, (b) $-\varepsilon_K$, and (c) $|\psi|$ for $L1_0$-MnAl films. The error bars correspond to the mean square root of the statistical errors, which are below 3.5 % for the values of $\theta_K$, $-\varepsilon_K$, and $|\psi|$.

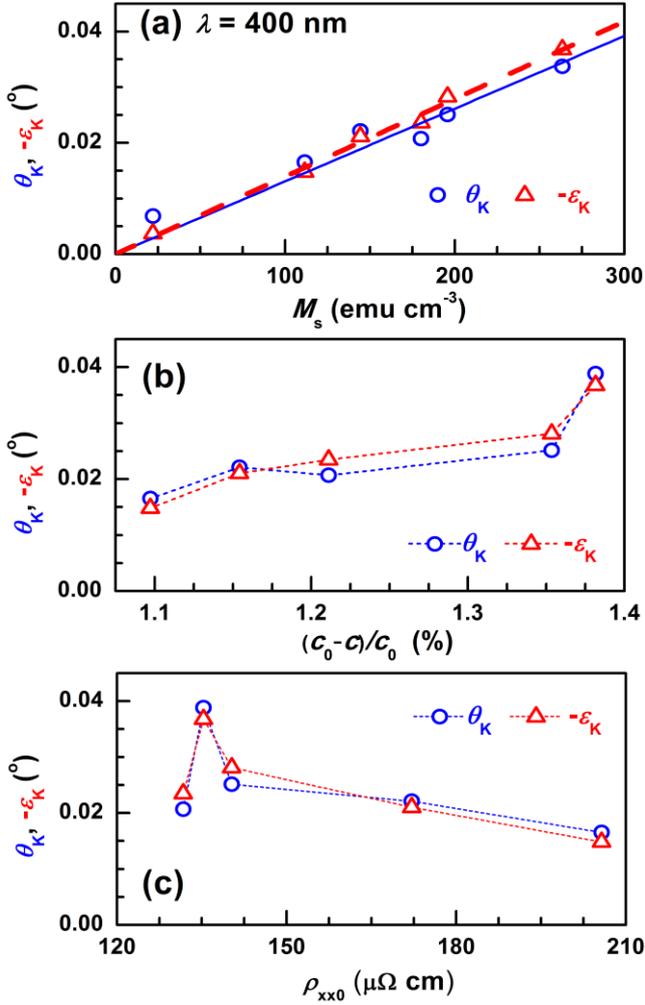

FIG. 4 $\theta_K$ and $-\varepsilon_K$ at $\lambda$ = 400 nm plotted as a function of (a) $M_s$, (b) $(c_0 - c)/c_0$, and (c) $\rho_{xx0}$ for $L1_0$-MnAl films with different $T_s$. In (a), the blue solid (red dashed) line represents the best linear fit for $\theta_K$ ($-\varepsilon_K$).

TABLE I. The saturation magnetization ($M_s$), the out-of-plane lattice constant ($c$), $(c_0-c)/c_0$, the residual resistivity ($\rho_{xx0}$), and perpendicular magnetic anisotropy ($K_u$) at room temperature for $L1_0$-MnAl films with different $T_s$. The out-of-plane lattice constant of $L1_0$-MnAl bulk $c_0$ is 3.570 Å [34]. The values of $M_s$, $c$, $K_u$, and $\rho_{xx0}$ were taken from Ref. [18] and Ref. [20,21], respectively. The uncertainty is < 0.18%, < 0.18%, ~0.01% and < 4 % for $c$, $(c_0-c)/c_0$, $\rho_{xx0}$ and $K_u$, respectively.

| $T_s$ (°C) | $M_s$ (emu cm$^{-3}$) | $c$ (Å) | $(c_0-c)/c_0$ (%) | $\rho_{xx0}$ (μΩ cm) | $K_u$ (Merg cm$^{-3}$) |
|---|---|---|---|---|---|
| 150 | 22.1 ±0.5 | - | - | - | 0.004 |
| 200 | 111.6 ±1.0 | 3.501 | 1.10 | 205.7 | 0.143 |
| 250 | 144.4 ±0.4 | 3.499 | 1.15 | 172.1 | 1.640 |
| 300 | 180.1 ±1.2 | 3.497 | 1.21 | 131.8 | 3.745 |
| 350 | 263.4 ±1.4 | 3.491 | 1.38 | 135.3 | 7.735 |
| 400 | 195.7 ±0.9 | 3.492 | 1.35 | 140.3 | 4.760 |

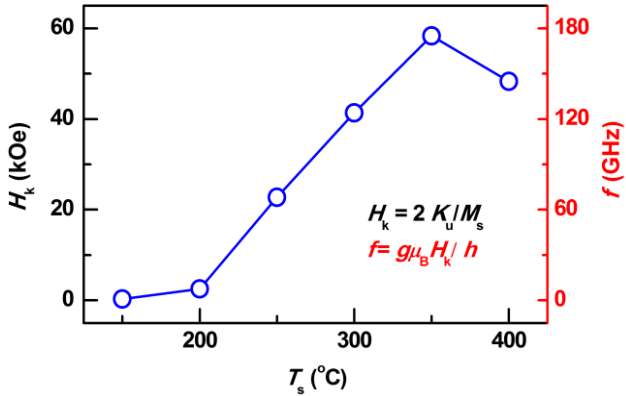

FIG. 5 The uniaxial anisotropic fields ($H_k$) and the estimated spin precession frequencies ($f$) of $L1_0$-MnAl films with different $T_s$.